\documentstyle{article}
\begin{document}

%                          Titel
\begin{center}
{\Large \bf Surface observables and the Weyl anomaly} \\

\vspace{4mm}

%                      author/address
M{\aa}ns Henningson\\
Institute of Theoretical Physics, Chalmers University of Technology\\
S-412 96  G\"oteborg, Sweden\\
\end{center}

%                       Abstract
\begin{abstract}
I review the computation of the conformal anomaly of a Wilson surface 
observable in free two-form gauge theory in six dimensions.
\end{abstract}

%                         Text

%Tables and figures with captions must be build into the text or
% placed after the bibliography. Can be used \psfig or \epsfig .

%                 Fig.1
%\begin{figure}[thb]
%\centerline{\psfig{figure=ptxda.eps,height=12cm,width=8cm}}
%\caption{\label{p_tscat} Scatter plots   of the combined CTD+VXD tracks:
%the left figure for tracks with $p_{t\,lab}>10$\,GeV/c, the right figure
%for  tracks from the region (18) with $p_{t\,lab}>100$\, GeV/c.}
%\end{figure}

%                         Text
$M$-theory admits configurations with some number $N$ of coincident five-branes embedded in a curved eleven-dimensional space. Small fluctuations around this configuration are described by a $(0, 2)$ superconformal field theory defined on the six-dimensional world-volume $M$ of the branes, and with a background metric $G_{\mu \nu}$ given by the embedding. There is a conformally invariant observable $W (\Sigma)$ associated with every two-cycle $\Sigma$ in $M$. When $N = 1$, this theory is simply given by a free $(0, 2)$ tensor multiplet, i.e. a two-form $B_{\mu \nu}$ with self-dual field strength, eight spinors $\psi^i$ and five scalars $\phi^a$, and the observable associated with a two-cycle $\Sigma$ is given by the exponential of the integral of $B_{\mu \nu}$ over $\Sigma$:
\begin{equation}
\left<W (\Sigma) \right> = \left<\exp 2 \pi i \int_\Sigma B \right> .
\end{equation}
For finite $N$, the theory is a difficult to study interacting superconformal field theory, about which rather little is known. For large $N$, however, there is a `holographic' description by a weakly coupled supergravity theory on an open seven-manifold $X$ such that $\partial X = M$. The expectation value of the observable associated with a two-cycle $\Sigma$ is then given by the exponential of the volume of an embedded minimal open three-manifold $D$ such that $\partial D = \Sigma$:
\begin{equation}
\left<W (\Sigma) \right> = \exp 2 \pi i \int_D V_D ,
\end{equation}
where $V_D$ is the volume form on $D$.

The theory itself, and also the observable $W (\Sigma)$ for an arbitrary two-cycle $\Sigma$, are invariant under conformal rescalings of the background metric $G_{\mu \nu}$. Formally, the quantity
\begin{equation}
I = - \frac{1}{4 \pi^2} \log \left<W (\Sigma) \right>
\end{equation}
is therefore conformally invariant. However, $I$ diverges and needs to be regularized. By adding local counterterms, we can cancel the divergences and obtain a finite $I_{\rm fin}$. This procedure can be carried out in a manifestly covariant way, but conformal invariance is in general lost. Under a conformal transformation 
\begin{equation}
\delta G_{\mu \nu} = 2 \phi G_{\mu \nu}
\end{equation}
with infinitesimal parameter function $\phi$,  $I_{\rm fin}$ transforms as
\begin{equation}
\delta I_{\rm fin} = - 4 \pi^2 \int_\Sigma d^2 \sigma \sqrt{g} {\cal A} ,
\end{equation}
where $g$ is the determinant of the induced metric $g_{\alpha \beta}$ on $\Sigma$, and the conformal or Weyl anomaly ${\cal A}$ is given by some local expression linear in $\phi$.

By the Wess-Zumino consistency conditions, the conformal variation of the anomaly ${\cal A}$ is a total derivative. It follows that ${\cal A}$ must be a linear combination of the terms
\begin{eqnarray}
& & R_{(2)} \phi \cr
& & \left(G_{\mu \nu} \nabla^2 X^\mu \nabla^2 X^\nu - 4 g^{\alpha \beta} P_{\alpha \beta} \right) \phi \cr
& & g^{\alpha \gamma} g^{\beta \delta} W_{\alpha \beta \gamma \delta} \phi \cr
& & \nabla^2 X^\mu D_\mu \phi , 
\end{eqnarray}
where $R_{(2)}$ is the curvature scalar of the induced metric $g_{\alpha \beta}$ on $\Sigma$, $\nabla^2 X^\mu$ is the mean curvature vector of the embedding of $\Sigma$ in $M$, and $P_{\alpha \beta}$ and $W_{\alpha \beta \gamma \delta}$ are the pullbacks of the tensor 
\begin{equation}
P_{\mu \nu} = \frac{1}{4} \left(R_{\mu \nu} - \frac{1}{10} R G_{\mu \nu} \right)
\end{equation}
and the Weyl tensor $W_{\mu \nu \rho \sigma}$ respectively to $\Sigma$.

For large $N$, the anomaly ${\cal A}$ was calculated in \cite{Graham-Witten} by generalizing the computation in \cite{Henningson-Skenderis98} with the result
\begin{equation}
{\cal A} \sim \left(G_{\mu \nu} \nabla^2 X^\mu \nabla^2 X^\nu - 4 g^{\alpha \beta} P_{\alpha \beta} \right) \phi - 2 \nabla^2 X^\mu D_\mu \phi .
\end{equation}
In this talk, I will review the calculation of ${\cal A}$ for $N = 1$ but without any self-duality condition imposed on the field strength of $B_{\mu \nu}$. For more details, I refer to our paper \cite{Henningson-Skenderis99}.

The gauge fixed action for $B_{\mu \nu}$ is
\begin{equation}
S = \int_M d^6 x \sqrt{G} G^{\lambda \rho} G^{\mu \sigma} G^{\nu \tau} \left( - \frac{3}{4} D_{[\lambda} B_{\mu \nu]} D_\rho B_{\sigma \tau} - \frac{\alpha}{2} D_\lambda B_{\rho \mu} D_\nu B_{\tau \sigma} \right)
\end{equation}
with an arbitrary non-zero constant $\alpha$ multiplying the gauge fixing term. Apart from this term, the action is conformally invariant, which means that the conformal variation of the propagator $\Delta_{\mu \nu ; \rho \sigma} (X, X^\prime)$ is a total derivative. The same is true for its pullback $\Delta_{\alpha \beta ; \gamma \delta} (\sigma, \hat{\sigma})$ to an embedded two-cycle $\Sigma$, i.e. 
\begin{equation}
\delta \Delta_{\alpha \beta ; \gamma \delta} (\sigma, \hat{\sigma}) = \partial_{[\alpha} \Lambda_{\beta] ; \gamma \delta} (\sigma, \hat{\sigma}) + \hat{\partial}_{[\gamma} \Lambda^\prime_{|\alpha \beta| ; \delta]} (\sigma, \hat{\sigma})
\end{equation}
for some $\Lambda_{\beta; \gamma \delta} (\sigma, \hat{\sigma})$ and $\Lambda^\prime_{\alpha \beta; \delta} (\sigma, \hat{\sigma})$ that are linear in the parameter $\phi$ of a conformal variation of the metric. It follows that the quantity
\begin{equation}
I = \int_\Sigma d \sigma^\alpha \wedge d \sigma^\beta \int_\Sigma d \hat{\sigma}^\gamma \wedge d \hat{\sigma}^\delta \Delta_{\alpha \beta; \gamma \delta} (\sigma, \hat{\sigma})
\end{equation}
is formally invariant under a conformal transformation.

However, $\Delta_{\alpha \beta; \gamma \delta} (\sigma, \hat{\sigma})$ diverges along the diagonal of $\Sigma \times \Sigma$, so $I$ is not really well defined. We therefore consider the regulated quantity
\begin{equation}
I_\epsilon  = \int_\Sigma d \sigma^\alpha \wedge d \sigma^\beta \int_\Sigma d \hat{\sigma}^\gamma \wedge d \hat{\sigma}^\delta \Theta(s^2 (\sigma, \hat{\sigma}) - \epsilon^2) \Delta_{\alpha \beta; \gamma \delta} (\sigma, \hat{\sigma}) ,
\end{equation}
where $\Theta( t)$ is the step function, $s^2 (\sigma, \hat{\sigma})$ is the square of the geodesic distance and $\epsilon > 0$ is a cutoff. We expect that
\begin{equation}
I_\epsilon = \epsilon^{-2} I_2 + \log \epsilon \, I_0 + I_{\rm fin} + {\cal O} (\epsilon) ,
\end{equation}
where the coefficients $I_2$ and $I_0$ of the divergent terms are given by some local expressions that can be cancelled by counterterms so that we are left with the finite quantity $I_{\rm fin}$. This regularization breaks conformal invariance, and under a conformal variation we expect that
\begin{equation}
\delta I_\epsilon = \epsilon^{-2} {\cal A}_2 + {\cal A} + {\cal O} (\epsilon) ,
\end{equation}
where ${\cal A} = \delta I_{\rm fin}$ is the conformal anomaly. On the other hand, we have that
\begin{eqnarray}\label{deltaIepsilon}
\delta I_\epsilon & = & \int_\Sigma d \sigma^\alpha \wedge d \sigma^\beta \int_\Sigma d \hat{\sigma}^\gamma \wedge d \hat{\sigma}^\delta \delta (s^2 (\sigma, \hat{\sigma}) - \epsilon^2) \cr
& & \left(\delta s^2 (\sigma, \hat{\sigma}) \Delta_{\alpha \beta; \gamma \delta} (\sigma, \hat{\sigma}) - \partial_\alpha s^2 (\sigma, \hat{\sigma}) \Lambda_{\beta; \gamma \delta} (\sigma, \hat{\sigma}) - \hat{\partial}_\gamma s^2 (\sigma, \hat{\sigma}) \Lambda^\prime_{\alpha \beta; \delta} (\sigma, \hat{\sigma}) \right) ,
\end{eqnarray}
where $\delta s^2 (\sigma, \hat{\sigma})$ denotes the conformal variation of the geodesic distance.

We see from the above that $\delta I_\epsilon$, and thus the anomaly, can be calculated just from a knowledge of the short distance expansion of the propagator, the geodesic distance and their conformal variations. In Riemann normal coordinates around $X^\prime$ and with $X^\mu = X^{\prime \mu} + \Delta X^\mu$, these are given by
\begin{eqnarray}
\Delta_{\rho \sigma; \mu \nu} (X, X^\prime) & = & - \frac{1}{4 \pi^3} |\Delta X|^{-4} \left[ \eta_{\rho \mu} \eta_{\sigma \nu} - \frac{1}{3} \left( P_{\mu \kappa} \eta_{\lambda \rho} \eta_{\sigma \nu} + \eta_{\mu \kappa} P_{\lambda \rho} \eta_{\sigma \nu} \right) \Delta X^\kappa \Delta X^\lambda \right. \cr
& & \left. + \left(\frac{4}{3} P_{\rho \mu} \eta_{\sigma \nu} - \frac{1}{2} W_{\rho \mu \sigma \nu} \right) |\Delta X|^2 + {\cal O} (\Delta X^3) \right] \cr
\delta \Delta_{\rho \sigma; \mu \nu} (X, X^\prime) & = & - \frac{1}{4 \pi^3} \partial_\rho |\Delta X|^{-2} \left( \eta_{\sigma \nu} \partial_\mu \phi + \frac{1}{2} \Delta X^\tau \eta_{\sigma \nu} \partial_\mu \partial_\tau \phi + {\cal O} (\Delta X^2) \right) \cr
& & - \frac{1}{4 \pi^3} \partial^\prime_\mu |\Delta X|^{-2} \left(\eta_{\nu \sigma} \partial_\rho \phi + \frac{1}{2} \Delta X^\tau \eta_{\nu \sigma} \partial_\rho \partial_\tau \phi + {\cal O} (\Delta X^2) \right)
\end{eqnarray}
antisymmetrized in $\rho \sigma$ and $\mu \nu$ with the tensors evaluated at $X^\prime$, and
\begin{eqnarray}
s^2 (X, X^\prime) & = & |\Delta X|^2 + {\cal O} (\Delta X^5) \cr
\delta s^2 (X, X^\prime) & = & |\Delta X|^2 \left( 2 \phi + \partial_\mu \phi \Delta X^\mu + \frac{1}{3} \partial_\mu \partial_\nu \phi \Delta X^\mu \Delta X^\nu \right) + {\cal O} (\Delta X^5) .
\end{eqnarray}
respectively.

The computations are most easily performed by introducing Riemann normal coordinates (with respect to the induced metric) on $\Sigma$ as well. The delta function in (\ref{deltaIepsilon}) reduces the integral over $d^2 \hat{\sigma}$ to a one-dimensional integral, which is elementary. We are then left with an expression integrated over $d^2 \sigma$ only. The end result is that
\begin{equation}
\delta I_\epsilon = - \frac{1}{4 \pi^2} \int_\Sigma d^2 \sigma \sqrt{g} \left( \epsilon^{-2} 4 \phi + {\cal A} + {\cal O} (\epsilon) \right) ,
\end{equation}
where the anomaly ${\cal A}$ is given by
\begin{equation}
{\cal A} = \left(- \frac{3}{4} (\nabla^2 X^\mu \nabla^2 X^\nu G_{\mu \nu} - 4 g^{\alpha \beta} P_{\alpha \beta}) - \frac{1}{2} R_{(2)} - \frac{1}{6} g^{\alpha \gamma} g^{\beta \delta} W_{\alpha \beta \gamma \delta} \right) \phi - \frac{5}{6} \nabla^2 X^\mu D_\mu \phi .
\end{equation}
We note that the quadratic divergence indeed can be cancelled by a local counterterm, which in fact is just the area of $\Sigma$. There is also a logarithmic divergence in $I_\epsilon$ (but not in $\delta I_\epsilon$) which is likewise given by a local expression. Finally, we point out that ${\cal A}$ is {\it not} proportional to the large $N$ anomaly computed in \cite{Graham-Witten}.

The author is supported by the Swedish Natural Science Research Council (NFR).

%            REFERENCES

%            or here
%                     FIGURES
%\begin{figure}[tp]
%\centerline{\psfig{figure=ttph.eps,width=16cm}}
%\caption{\label{thetaphi}
%Distributions of tracks
%in polar and azimutal angles in ZS (the first row) and in BS (the second
%row). In the third row shown the  angle distributions  of
% electrons  detected  in  the DU callorimeter.
%}
%\end{figure}

\end{document}